\documentclass[default,iicol]{sn-jnl}

\usepackage{graphicx}%
\usepackage{multirow}%
\usepackage{amsmath,amssymb,amsfonts}%
\usepackage{amsthm}%
\usepackage{mathrsfs}%
\usepackage[title]{appendix}%
\usepackage{xcolor}%
\usepackage{textcomp}%
\usepackage{manyfoot}%
\usepackage{booktabs}%
\usepackage{algorithm}%
\usepackage{algorithmicx}%
\usepackage{algpseudocode}%
\usepackage{listings}%
\usepackage{natbib}
\usepackage{float}
\usepackage{caption}
\usepackage{placeins}
\setcitestyle{authoryear,open={(},close={)}}

\begin{document}

\title[Article Title]{Simulating Vision Impairment in Virtual Reality - A Comparison of Visual Task Performance with Real and Simulated Tunnel Vision}

\author*[1]{\fnm{Alexander} \sur{Neugebauer}}\email{a.neugebauer@uni-tuebingen.de}

\author[2]{\fnm{Nora} \sur{Castner}}\email{nora.castner@zeiss.com}

\author[1]{\fnm{Björn} \sur{Severitt}}\email{bjoern.severitt@uni-tuebingen.de}

\author[3]{\fnm{Katarina} \sur{Stingl}}\email{katarina.stingl@med.uni-tuebingen.de}

\author[2]{\fnm{Iliya} \sur{Ivanov}}\email{iliya.ivanov@zeiss.com}

\author[1,2]{\fnm{Siegfried} \sur{Wahl}}\email{siegfried.wahl@uni-tuebingen.de}

\affil[1]{\orgdiv{Institute for Ophthalmic Research, ZEISS Vision Science Lab}, \orgname{University of Tübingen}, \orgaddress{\street{Elfriede-Aulhorn-Str. 7}, \city{Tübingen}, \postcode{72076}, \country{Germany}}}

\affil[2]{\orgname{Carl Zeiss Vision International GmbH}, \orgaddress{\street{Turnstraße 27}, \city{Aalen}, \postcode{73430}, \country{Germany}}}

\affil[3]{\orgdiv{Center for Ophthalmology, University Eye Hospital}, \orgname{University of Tübingen}, \orgaddress{\street{Elfriede-Aulhorn-Str. 7}, \city{Tübingen}, \postcode{72076}, \country{Germany}}}

\abstract{\textbf{Purpose:} In this work, we explore the potential and limitations of simulating gaze-contingent tunnel vision conditions using Virtual Reality (VR) with built-in eye tracking technology. This approach promises an easy and accessible way of expanding study populations and test groups for visual training, visual aids, or accessibility evaluations. However, it is crucial to assess the validity and reliability of simulating these types of visual impairments and evaluate the extend to which participants with simulated tunnel vision can represent real patients.

\textbf{Methods:} Two age-matched participant groups were acquired: The first group (n=8, aged 20-60, average 49.1±13.2) consisted of patients diagnosed with Retinitis pigmentosa (RP). The second group (n=8, aged 27-59, average 46.5±10.8) consisted of visually healthy participants with simulated tunnel vision. Both groups carried out different visual tasks in a virtual environment for 30 minutes per day over the course of four weeks. Task performances as well as gaze characteristics were evaluated in both groups over the course of the study.

\textbf{Results:} Using the 'two one-sided tests for equivalence' method, the two groups were found to perform similar in all three visual tasks. Significant differences between groups were found in different aspects of their gaze behavior, though most of these aspects seem to converge over time.

\textbf{Conclusion:} Our study evaluates the potential and limitations of using Virtual Reality technology to simulate the effects of tunnel vision within controlled virtual environments. We find that the simulation accurately represents performance of RP patients in the context of group averages, but fails to fully replicate effects on gaze behavior.}


\maketitle

\clearpage

\section{Introduction}\label{sec1}

Virtual Reality (VR) has emerged as a powerful and increasingly popular tool in the field of vision research in recent years. The number of published research articles in the database of PubMed that incorporate the key words 'Virtual Reality' and 'Ophthalmology' have almost tripled in the time since 2019. The reasons for this growing interest and utilization of VR are readily apparent. VR offers a unique platform to investigate various aspects of human vision, perception, and gaze behavior, providing a multitude of advantages compared to other psychophysical experimental setups. Compared to computer-screen based setups \citep{hecht2015effects, ivanov2016eyeMovement}, VR setups provide larger viewing angles - typically between 90-100° \citep{sauer2022assessment}. They additionally allow to expand the displayed visual area even further by considering the head rotation of the user, potentially allowing for a full 360° visual scene. Both the stereoscopic view, provided by separate screens per eye, as well as the parallax effect that occurs when shifting the head allow for natural and realistic depth perception. With that, VR provides overall visual experiences closer to real life \citep{hibbard2023VR}. Meanwhile, the fully simulated nature of the environments within a VR-based setting allows for much higher flexibility and control compared to actual real-life setups. Experimental environments can be randomized or modified with the press of a button, and each detail and parameter can be individually adjusted without physical or material restraints. Furthermore, with the increasing availability of eye-tracking in VR-headsets \citep{adhanom2023eye}, the use of VR for studies on gaze behavior or simulation of specific visual conditions expands even further. 

One potential use for VR in vision research is the simulation of visual impairments \citep{Jones2018VRsimulation, hibbard2023VR, krosl2018VisionImpairmentEffects, krosl2019VRCataractSimulation, jones2020Perspectives}. An accurate simulation of visual impairments can have various fields of application. It could aid in the evaluation of accessibility of devices, software interfaces or environment architecture \citep{Jones2018VRsimulation, krosl2019VRCataractSimulation}. It could provide doctors, family members, caregivers, employers, or colleagues better insights into the experience with - and limitations caused by - visual impairments \citep{Jones2018VRsimulation, krosl2023EducationalValueOfVRSimulation, vayrynen2016VRCitySimulation}. Lastly, it could allow researchers and developers of visual aid systems and low vision training tools to expand a relatively small study population of real patients by including visually healthy participants with simulated visual conditions \citep{Acevedo2022LowVisionMixedReality}. This becomes especially relevant for rare conditions, as acquisition of sufficient study populations can be problematic in these cases \citep{mitani2020rareDiseases}.

Several studies have explored ways to design simulations for visual field defects to be as visually accurate as possible \citep{Geisler2002screenBasedSimulation, Lewis2011Ue3Simulation, Stock2018RealisticSimulation}, providing visual experiences similar to real conditions. However, visual representation is only part of the simulation. While VR provides a solid theoretical basis, there are many factors related to visual impairments that cannot be replicated even with the use of VR \citep{hibbard2023VR}. These factors include experience and adaptive behavior: Patients with visual impairments often develop specific behaviors over years to cope with their condition, whereas visually healthy participants using a simulation lack this adaptive experience. Moreover, cognitive distinctions exist between actual and simulated visual field defects. Individuals with these defects often initially lack awareness of their condition, especially in cases of gradual onset due to inherited retinal diseases \citep{Fletcher2012PatientAwareness,Hoste2003SubjectivePerception}. Even when aware of the diagnosis, affected patients typically do not describe the missing visual fields as actively perceived phenomena \citep{Hoste2003SubjectivePerception}; instead, they tend to recognize them only through a decline in everyday visual task performance. Consequently, representing visual field defects as black areas, which is a common practice when visualizing or simulating them using VR or other methods, will never fully capture the lived reality, as the participants will actively notice the visual limitations.

Acknowledging that simulations of visual field defects will - with current technological possibilities - never be perfect, it becomes essential to understand the extent to which these simulations can still accurately mirror real visual field impairments across various types of visual tasks and behaviors. Thus, this work aims to evaluate the performance and gaze behavior in participants with simulated tunnel vision in three different Virtual-Reality-based tasks. The results are compared to those of an age-matched group of actual patients with peripheral visual field defects (tunnel vision) caused by Retinitis pigmentosa (RP). The results of both groups are collected within the same experimental setup, allowing to evaluate the similarities and differences of task-specific performance parameters and gaze characteristics between the groups. To the best of our knowledge, this is the first study to directly investigate and compare results between patients with visual field defects and participants with simulations of these same defects. The findings of this work will provide an overview of the criteria that studies and test setups should consider when utilizing VR-simulated tunnel vision, and possibly provide a baseline for the evaluation of simulations of other types of visual field defects.

\section{Methods}

\subsection{Ethics}
This study was proposed to and approved by the ethics committee of the Institutional Review Board of the Medical Faculty of the University of Tübingen (628/2018BO2) in accordance with the 2013 Helsinki Declaration. All participants signed written informed consent forms. 

\subsection{Obtaining research data}
To evaluate and compare the performance and gaze behavior displayed by both real patients and participants with simulated conditions, two sets of data are required - one for the group of RP patients and one for the group of participants with simulated tunnel vision. In the following, the patient group will be called 'Group A', the group of participants with simulated tunnel vision will be referred to as 'Group B'.

The data for Group A originates from a study by Neugebauer et al. \citep{neugebauer2023gazeTraining} examining the effects of VR gaze training on visual performance and navigation performance of Retinitis pigmentosa patients. As part of this study, patients underwent four weeks of training with a Virtual-Reality based gaze training, consisting of different visual tasks, as will be described in section \ref{lab:VirtualRealitySetup}. The work by Neugebauer et al. mainly discusses the effects of the training on the performance and gaze behavior of patients in a real-world setting. However, eye-tracking data and data about the performance displayed by RP patients within the VR training were acquired as part of the results and thus provide a solid basis for the comparison targeted in this work. To acquire the second data set, a group of visually healthy participants (Group B) was acquired. To ensure compatibility and comparability between both data sets, participants of Group B were age-matched to the existing dataset of Group A. 

\subsection{Study population} \label{lab:studyPopulation}
Group A consists of eight RP patients (1 male, 7 female) with ages ranging from 20 to 60 years (average 49.6 ± 13.0) and VF sizes of 7° - 25° diameter. Group B comprises of eight participants with healthy or corrected vision (3 male, 5 female), aged between 27 and 59 (average 46.5 ± 10.8). The two groups were matched based on their age and their experience with VR devices, with a standard deviation of ±8.4 years between the two groups. 

Table \ref{tab:patient_data} lists information about the participants of both groups. Information on the patients' age of diagnosis, their level of experience with VR, and their used vision correction are based on participants' reports. The visual acuity of patients was provided based on their most recent medical examination. It was ensured at the start of the training that all participants were able to effortlessly recognize all relevant elements within the virtual environment. The reported visual fields (VFs) were measured independently, as described in section \ref{lab:MeasurementAndSimulation}. However, the VFs as reported in the official medical examination can be found in Appendix A. 

\begin{table*}[htbp]
\begin{center}
\caption{List of participant data, matched pairs being indicated by the same number. For Group B, the Visual Field Diameter describes the simulated VF.}
\begin{tabular}{|p{1.6cm}||p{1.4cm}|p{3.1cm}|p{2.2cm}|p{1.5cm}|p{1.7cm}|p{0.9cm}|l|}
\hline
Participant (Age\textbar Sex) & Age of diagnosis & {Visual Field\newline Diameter (RE / LE)} & {Visual Acuity (RE / LE)} & VF notes & VR\newline Experience & Vis. Cor.\footnotemark[1]\\
\hline
 \multicolumn{7}{|c|}{Group A - Retinitis pigmentosa patients} \\
\hline
1 (20f) & 14 & 7.62° / 8.26° & 0.40 / 0.40 & - & high & G/C\\
2 (57f) & 27 & 18.64° / 18.18° & 0.20 / 0.05 & spots\footnotemark[2] & - & G \\
3 (55f) & 18 & 17.64° / 16.36° & 0.13 / 0.20 & - & - & G \\
4 (47m) & 25 & 24.60° / 25.40° & 0.05 / 0.05 & - & low & G \\
5 (59f) & 50 & 18.54° / 18.34° & 0.32 / 0.25 & spots\footnotemark[2] & - & G \\
6 (59f) & 16 & 10.92° / 9.64° & 0.10 / 0.10 & - & - & G\\
7 (40f) & 18 & 12.18° / 14.56° & 0.40 / 0.32 & spots\footnotemark[2] & - & G \\
8 (60f) & 20 & 20.00° / 19.48° & 0.50 / 0.40 & - & - & G\\
\hline
 \multicolumn{7}{|c|}{Group B - Participants with simulated tunnel vision} \\
\hline
1b (27m) & - & 7.62° / 8.26° & - & - & med. & G\\
2b (37m) & - & 18.64° / 18.18° & - & - & low & -\\
3b (55f) & - & 17.64° / 16.36° & - & - & - & G\footnotemark[3]\\
4b (37f) & - & 24.60° / 25.40° & - & - & - & -\\
5b (56m) & - & 18.54° / 18.34° & - & - & - & - \\
6b (59f) & - & 10.92° / 9.64° & - & - & - & G\\
7b (47f) & - & 12.18° / 14.56° & - & - & - & G\footnotemark[3]\\
8b (54f) & - & 20.00° / 19.48° & - & - & - & G\\
\hline
\end{tabular}
\label{tab:patient_data}
\end{center}
\vspace{2mm}
\textsuperscript{1}Vision  correction used by the participant; G: Glasses; C: Contact lenses.\\
\textsuperscript{2}The patient displays some spots of remaining vision in the peripheral field.\\
\textsuperscript{3}Did not wear glasses during VR training.\\

\footnotetext[1]{Vision  correction used by the participant; G: Glasses; C: Contact lenses.}
\footnotetext[2]{The patient displays some spots of remaining vision in the peripheral field.}
\footnotetext[3]{Did not wear glasses during VR training.}
\end{table*}

\subsection{Experimental design}
In this section, we will describe the Virtual Reality software applied to test the performance, learning rate, and gaze characteristics of participants. The setup - both hardware and software - used for Group B within this study is identical to that used for Group A in the preliminary study. The only difference is the addition of a tunnel vision simulation in the setup for Group B, as will be elaborated further in section \ref{lab:MeasurementAndSimulation}

\subsubsection{Software and hardware specifications} \label{lab:specs}
The Virtual-Reality environment and the visual tasks applied in this study were developed with the Unity 3D game engine (Version 2021.3LTS) using the  Pico XR SDK (Version 1.2.4). The software was installed on the Pico Neo 2 Eye Virtual Reality headset, which features an 89° Field of View according to Sauer et al. \citep{sauer2022assessment} at a 75Hz refresh rate.

The Pico Neo 2 Eye uses the tobii eye tracking system with 90Hz frequency and 0.5° accuracy, according to official technical specifications \citep{tobii2023website}. Stein et al. \citep{Stein2021Latency} determined the latency of eye-tracking for this system to be 50ms, with an additional 29ms display latency, for an end-to-end latency of 79ms. 

\subsubsection{Visual field measurement and implementation of simulated visual field defects} \label{lab:MeasurementAndSimulation}
To achieve comparable conditions for both groups, the tunnel vision simulation applied for Group B was directly based on the actual VF size of the age-matched patient from Group A. While VF data was available in the medical examination reports provided by patients in Group A, it was determined that conducting a separate VF test within the virtual environment used for the study would be more practical for simulation purposes. This decision was motivated by two key factors. Firstly, for comparability it is essential to have a uniform VF measurement with standardized settings for all patients. The VF reports in the medical examinations came from different examiners and were conducted using varying setups, with no consistent reporting of the specific testing methods and perimetry settings. Additionally, visual angles are difficult to align between virtual environment and real world, as they are influenced by the distance between the user's eyes and the lenses and screens of the VR device. Therefore, to achieve an accurate simulation of the patients' vision within the virtual environment, it is recommended to measure the VF directly within this virtual setting.

To achieve this goal, a customized kinetic perimetry test was developed. In this test, a white target with a radius of 0.72° is presented on a black background. The target starts at the outer periphery of the VF at 45° and moves toward the center at a constant speed of 3° per second. In the center of the VF, a marker is displayed, and patients are instructed to focus their gaze on this central marker. During the perimetry process, the direction of the patients' gaze is continuously monitored. If the gaze deviates from the central marker, the moving target disappears. This precautionary measure is taken to prevent patients from unintentionally or intentionally exploring the scene in search of the target, which could compromise the accuracy of the results. The perimetry test was conducted at the maximum display brightness available on the Pico Neo 2 Eye, which, as noted in Sauer et al. \citep{sauer2022assessment}, corresponds to a luminance of $60\frac{cd}{m^2}$. The angular resolution of the perimetry test was set at 15°, resulting in the acquisition of a total of 24 data points per trial. Each eye was tested individually by displaying the target on one screen at a time, making it visible in only one eye. The results of this perimetry test and a comparison with the VF sizes reported in the patients' most recent medical diagnoses are presented in Appendix A. It is important to note that this perimetry test is not diagnostically validated. However, it provides a close estimation of the dimensions of the patients' VFs within the VR setup. As such, these results serve as a suitable parameter for simulating the corresponding visual field defects.

Based on the measured VF dimensions, alpha masks were created using the image editing software Gimp (version 2.10). The peripheral regions of this mask were colored in black, with the VF area being left transparent. The edges of the VF mask were blurred, resulting in a gradual transition from transparent to black over an edge of $\sim$2° visual angle. The alpha masks were then implemented in the VR software setups for Group B, using eye tracking data such that the masks move in a gaze-contingent manner.

\subsubsection{Virtual Reality setup} \label{lab:VirtualRealitySetup}
The VR software applied both in the preliminary study \citep{neugebauer2023gazeTraining} and in this study consists of three different visual tasks. Each task focuses on a different issue related to the lack of peripheral vision: Motion perception and tracking, visual search, and navigation. A short video demonstrating the three tasks is found in the supplementary files (Video S1).

\begin{figure*}[htbp]%
\centering
\includegraphics[width=\textwidth]{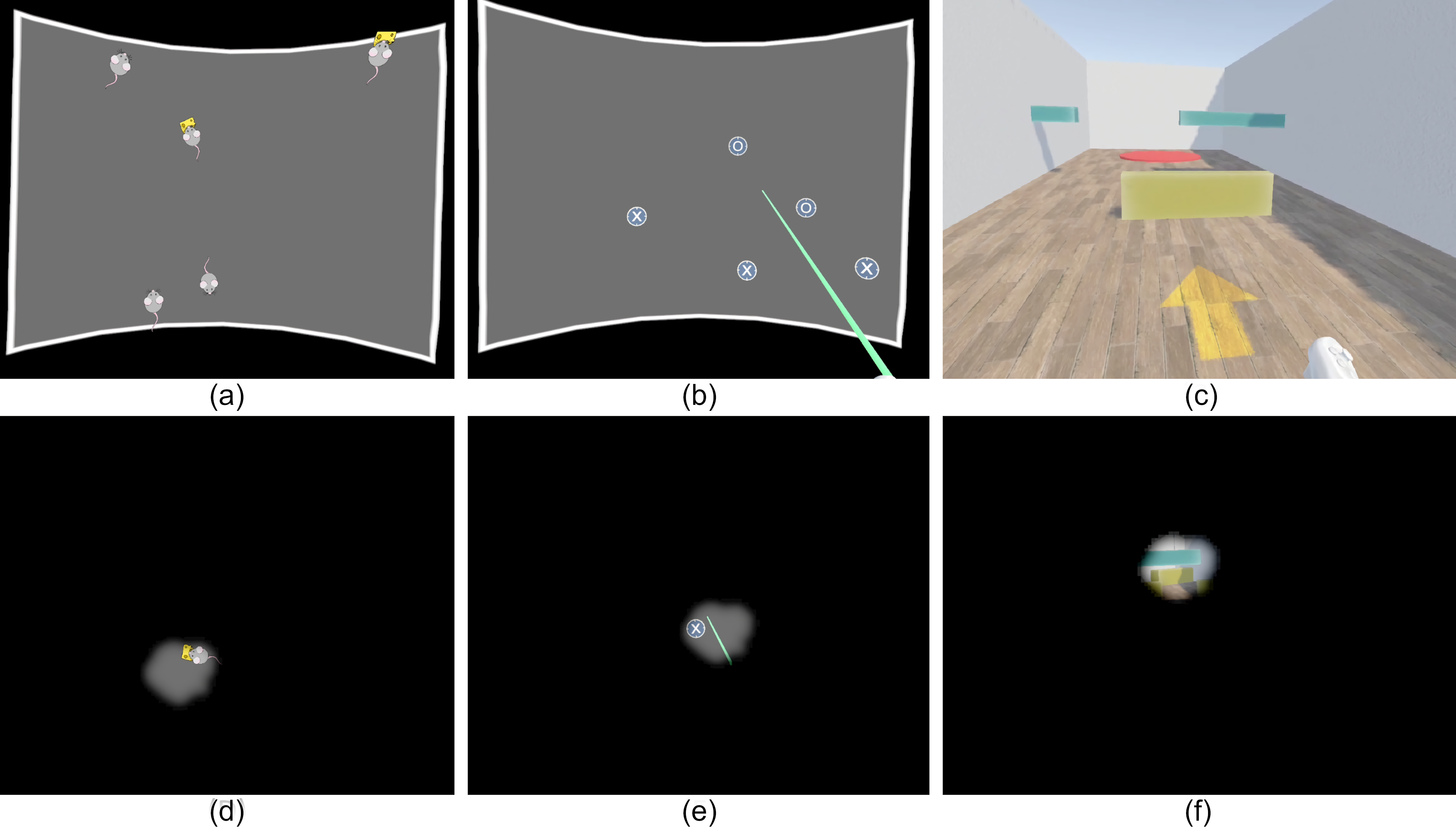}
\caption{Examples of the three visual tasks carried out by the participants, from \citep{neugebauer2023gazeTraining}. (a-c) show the tasks without tunnel vision simulation. (d-f) show the tasks with an exemplary simulated deficient visual field of $\sim$15° diameter, giving an impression of the visuals displayed to the participant group with simulated tunnel vision. (a,d) Target Tracking; (b,e) Search task; (c,f) Navigation task. The interactive area in the Target Tracking and Search task (visbile in a and b) have dimensions of 80° horizontally and 60° vertically. }
\label{fig:VisionTasks}
\end{figure*}

\textbf{Target tracking} The peripheral VF is a crucial component in motion perception \citep{finlay1982motion}. Not only does it allow for the initial perception of movement outside of the foveated area; it also enables a person to consistently track movement and estimate motion paths, thus determining potential risks and the danger of collision. A lack of the peripheral VF greatly impairs this ability. While a single moving object can be tracked by keeping the gaze directed at it, tracking multiple moving objects within a scene simultaneously becomes impossible without switching focus between them. The target tracking task (Fig. \ref{fig:VisionTasks} a and d) aims to test and improve the participants' ability to estimate motion paths of moving targets and to frequently, accurately, and quickly switch focus between targets. In the beginning of the task, a number of identical-looking targets (visualized as mice for higher visual appeal) are spawned at the center of a defined 'play area'. A subset of these targets is marked by a visually distinct indicator (a piece of cheese) (Fig. \ref{fig:VisionTasks} a). During the trial, all targets move randomly across the play area at a fixed speed, changing directions over time, and the participant is tasked to track all marked targets. In order to prevent one target being occluded by another, targets adjust their direction when getting close to a different target, avoiding collisions. Similarly, when targets approach the boundary of the play area, they adjust their movement towards a point within the area. After a random time interval of 8-12 seconds, all targets stop their movement and the visual markers disappear, rendering all targets in the area visually identical. At this point, the participant is tasked to select all previously marked targets. Based on the number of incorrectly selected targets, the task is rated as full success (zero incorrect targets), half success (one incorrect target) or failure (two or more incorrect targets). The number of marked and unmarked targets, the movement speed of targets as well as the dimensions of the play area are all based on a variable difficulty level. The number of targets starts at five, with two marked targets. The VF dimensions start at 52° horizontally and 39° vertically (approximately 30\% of the dimensions of a healthy VF) and the movement speed of mice starts at 3° per second. If the participant's performance reaches a specified threshold, the difficulty level will increase, which leads to higher numbers of targets, higher target movement speed and larger play area. If the participant performs below the threshold, it is also possible that the difficulty level is reduced. This paradigm ensures that the task remains challenging even when participants become more experienced with it. Difficulty levels range from 1 to 100, with each level increasing the adjustable parameters by approximately 3-5\% of the starting value. 

\textbf{Search task} Visual search is another aspect greatly influenced by the peripheral VF \citep{David2021search}. Visual stimuli from the periphery inform us about potential regions of interest, which incentivizes gaze movements towards these areas. While a lack of the peripheral field does not impact the ability to recognize search targets as soon as they enter the VF, it prevents visual stimuli that would guide the gaze towards areas where the target might be located. Thus, people living with severe tunnel vision depend on a more consciously controlled gaze movement to 'scan' their surroundings for any visual cue or target. The search task, based on a similar approach by Ivanov et al. \citep{ivanov2016eyeMovement}, aims to test and improve this consciously controlled gaze movement. A number of targets (visible in Fig. \ref{fig:VisionTasks} b and e) with a radius of 2° visual angle are spawned within the play area. Three targets are marked visually distinct from the others through a large X symbol in the center of the target. Participants are tasked to scan the play area over the course of a 20 second time window in search for these targets marked with an X, selecting them when found. Meanwhile, distractor targets are marked with an O and are not to be selected. Once a target is selected, it is removed and a new target is generated within the play area, but outside of the participant's VF. This ensures that there are always exactly three marked targets within the play area. The goal of the task is to find and select as many marked targets as possible over the duration. Similar to the tracking task, the difficulty of the search task is adaptable, increasing or decreasing in difficulty gradually based on the participant's performance. A higher difficulty level results in a larger play area and a higher ratio of distractor targets.

\begin{figure}[htbp]%
\centering
\includegraphics[width=\columnwidth]{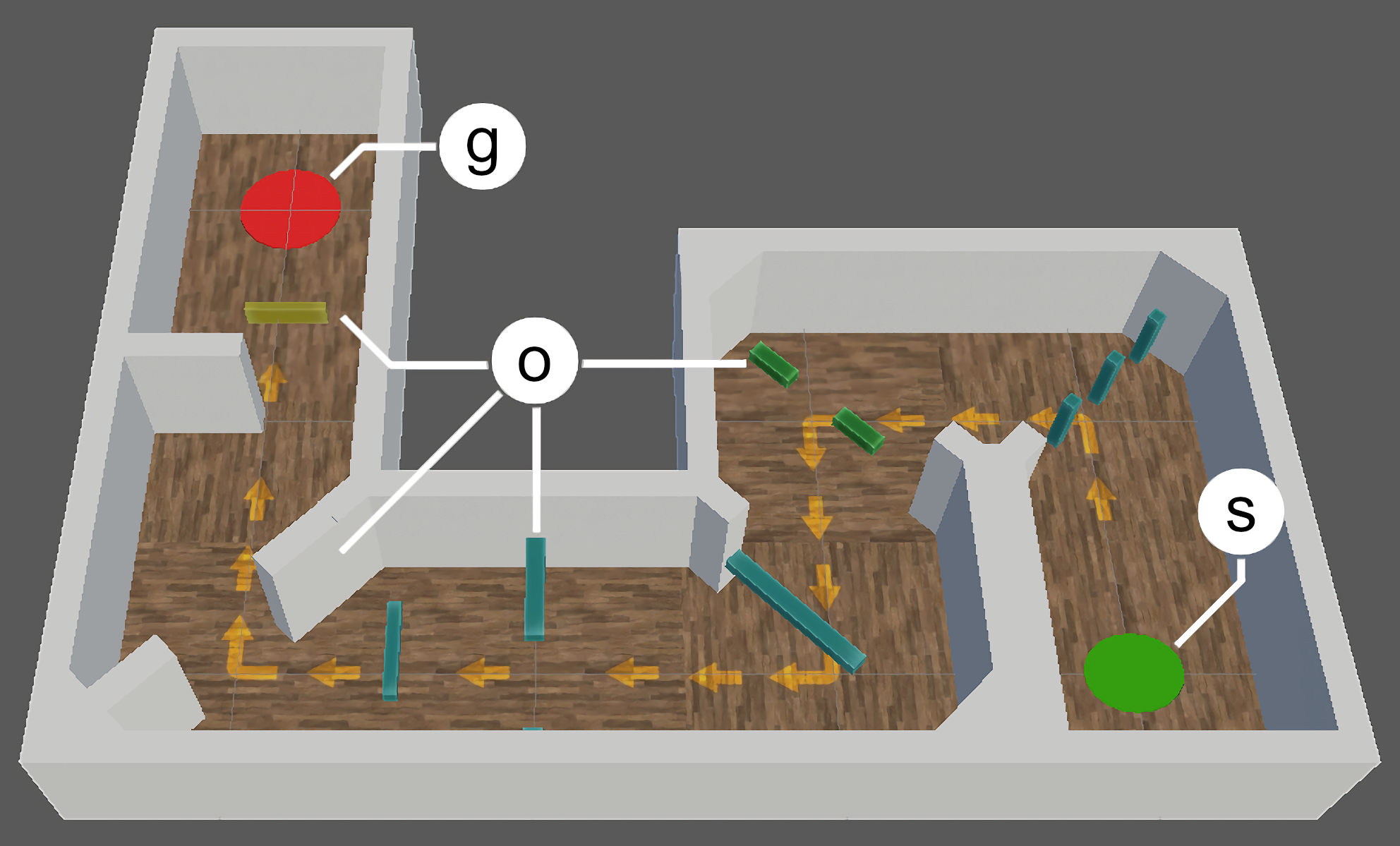}
\caption{Example image of one of the obstacle course layouts used in the navigation task, from \citep{neugebauer2023gazeTraining}. s: Starting position; o: Obstacles (example selection); g: Goal area.}
\label{fig:VRParkourTopDown}
\end{figure}

\textbf{Navigation task} A third task known to often cause difficulties for people with peripheral VF loss is navigation \citep{barhorst2016navigation}. Severe VF restrictions not only increase the risk of collision due to not spotting an obstacle in time, it can also impair efficient movement path planning and spatial memory \citep{barhorst2016navigation}. Thus, the third task of the gaze training (Fig. \ref{fig:VisionTasks} c and f) focuses on testing and improving the navigation and obstacle awareness of participants. For each trial, a randomized obstacle course is generated within the virtual environment (Fig. \ref{fig:VRParkourTopDown}). 

The course consists of an 8m wide and approximately 56m long corridor. Each corridor features two straight tiles, two right corners and two left corners, all in randomized order, for a total of 90 unique layouts. Within the course, a randomized selection of 12 different obstacles are placed. These obstacles consist of walls, near-ground obstacles, obstacles hanging from the ceiling, and obstacles moving from one side of the corridor to the other in a repetitive motion. The participant is tasked to navigate through this obstacle course using their body orientation for direction and controller input for walking speed. Their goal is to reach a zone at the end of the obstacle course (Fig. \ref{fig:VRParkourTopDown} g) while avoiding collisions, and in as short an amount of time as possible. The different types of obstacles encourage varying gaze behavior, such as scanning both on ground level as well as on head level. A collision is indicated to the participant through an audio cue. A higher difficulty level reduces the time available to participants to move through the obstacle course, though it also increases the movement speed of the avatar. Thus, in order to perform well on higher difficulty levels, participants are required to move and react faster to avoid obstacles.

\vspace{3mm}

\noindent Once a trial is completed, participants are brought back into a menu screen in which they receive information on their performance within the trial, as well as information on their overall progress, such as the number of daily completed trials and the current difficulty level for the task. Participants were able to mark a trial as 'invalid' using a specified option in this menu. This option was implemented for cases where technical difficulties or outside distractions interrupted the trial execution.

\subsection{Study layout}
\begin{figure*}[htbp]%
\centering
\includegraphics[width=\textwidth]{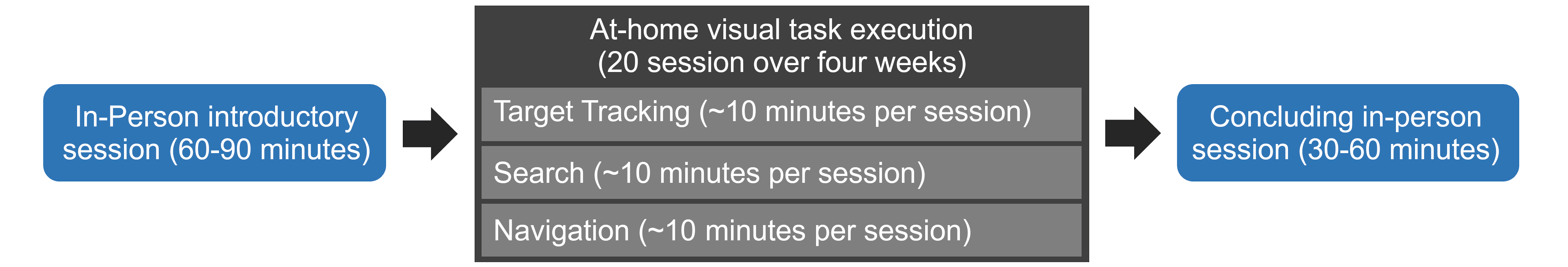}
\caption{Schematic of the study layout, showing the introductory in-person session, the task execution phase of 20 sessions of 30 minutes each over four weeks, and a concluding in-person session.}
\label{fig:ComparisonStudyLayout}
\end{figure*}

For both groups, the study layout consists of three sections (Fig. \ref{fig:ComparisonStudyLayout}): It starts with an introductory session in which the participants are introduced to the VR headset and shown how to operate within the virtual reality environment. The built-in eye tracking device was calibrated using the pre-implemented tobii eye tracking calibration tool, and after successful calibration, the task execution software was started. Participants went through a guided task execution session, carrying out all tasks for around 5-10 minutes. The experimenter explained the goals and controls for the three visual tasks as well as general menu controls. In addition, a gaze movement pattern was visualized and suggested to participants. This pattern describes a boustrophedon movement - a reading-like motion where the horizontal direction of movement alternates between left-right and right-left - and was based on findings of a previous study \citep{neugebauer2021scanningPattern} in which the pattern was found to have a mostly positive effect on visual performance.

Following that, participants were provided with a VR device and would execute the three visual tasks over the course of four weeks in an unsupervised at-home setting. Over the duration, a total of 20 sessions of 30 minutes each were carried out. Performance and eye tracking data were automatically tracked and stored on the device. In the concluding in-person session at the end of the task execution phase, participants returned the VR device to the experimenter.

\subsubsection{Questionnaire} During the task execution phase, participants were asked to fill out questionnaires to assess subjective ratings of enjoyment, motivation, stress, eye strain, and other factors of the task execution. The questionnaires had the format of a 10-point Likert scale \citep{sullivan2013Likert} and were filled after the first training session and after every five training sessions, for a total of 5 questionnaires filled over the course of the training. All five questionnaires featured the same seven questions:
\vspace{3mm}

\begin{itemize}
    \item \textbf{Enjoyment} To what extent do you find each of the visual tasks enjoyable?
    \item \textbf{Motivation} How motivated are you to improve your performance in each of the visual tasks?
    \item \textbf{Easiness} How would you rate the ease of carrying out each of the visual tasks?
    \item \textbf{Stress} To what degree do you experience stress while executing each of the visual tasks?
    \item \textbf{Eye Strain} How straining is each visual task on your eyes?
    \item \textbf{Intuitiveness} How intuitive is the use of the gaze training software?
    \item \textbf{Discomfort} How much physical discomfort do you experience while wearing the VR headset?
\end{itemize}

\subsection{Measurement parameters}
During training, different parameters were measured in both groups in order to evaluate the participants' performance in the individual visual tasks as well as their gaze behavior during these tasks. The parameters were stored on the VR headset and were evaluated after the training was finished. Parts of the results of the RP patient group have been reported before \citep{neugebauer2023gazeTraining}. 

\subsubsection{Performance parameters}
\begin{itemize}
    \item \textbf{Target tracking task performance} The target tracking task required participants to select previously marked targets at the end of a trial. The number of incorrectly selected targets - i.e. targets that were not marked during the trial - was measured and used to determine a score for this trial. If no incorrect target was selected before all correct targets are found, the trial was rated with a score of 2. If only one incorrect target was selected, the trial was rated with a score of 1. All other trials were rated with a score of 0.

    \item \textbf{Search task performance} In the search task, performance was measured by the number of static targets found and selected during the trial duration of 20 seconds.

    \item \textbf{Navigation task performance} In the navigation task, both trial duration and number of collisions were measured and individually assessed. Trial duration was defined as the duration from the start of a trial until the defined goal area was reached. Collisions were detected when a bounding capsule that was vertically attached to the user's avatar collided with any surface in the scene, either obstacles or the walls bounding the course. The capsule collider had a diameter of 0.4m and was always positioned exactly at the x- and y-coordinates of the scene camera, oriented vertical to the ground to simulate the body of a standing/walking human. 
\end{itemize}

When assessing performance using these parameters, the variable difficulty level of the tasks introduced an extra layer of complexity in evaluating participants' performance. As the difficulty increased, it was natural for participants to exhibit a decrease in absolute performance, as more challenging tasks inherently resulted in a higher number of failures. To address this challenge, additional measures were implemented. In the target tracking task, the difficulty level itself proved to be a suitable measure of patient performance. Due to the gradual increase or decrease of the difficulty level based on participants' performance, the difficulty level will usually converge towards a 'saturation' at which the participant will neither win nor lose a predominant number of trials. This saturated difficulty level can be seen as indicator for the current performance of the participant. However, in the two other tasks, this method was not feasible, as in both cases difficulty levels did not reach natural saturation for all participants. This was partly caused by participants improving at a constant rate over the entire duration of the study, or by participants reaching the lower limit of the difficulty range. Thus, for the search task, the performance score was instead adjusted in relation to the size of the play area in which targets were placed. This adjustment was calculated as follows: $P_{adj} = t*(w_{area} * h_{area})$, where $P_{adj}$ represents the adjusted performance score, $t$ denotes the number of targets found, and $w_{area}$ and $h_{area}$ represent the width and height (measured in visual angles) of the play area, respectively. This way, the increasing size of the play area and the resulting increased challenge to find targets within the area is considered in the score. For the navigation task, no specific measures were taken to adjust for changes in difficulty, as changes in difficulty had only minor effects on the measured parameters, and trials of different difficulty levels still remained largely comparable to each other.

\subsubsection{Gaze parameters}

\begin{figure*}[htbp]%
\centering
\includegraphics[width=\textwidth]{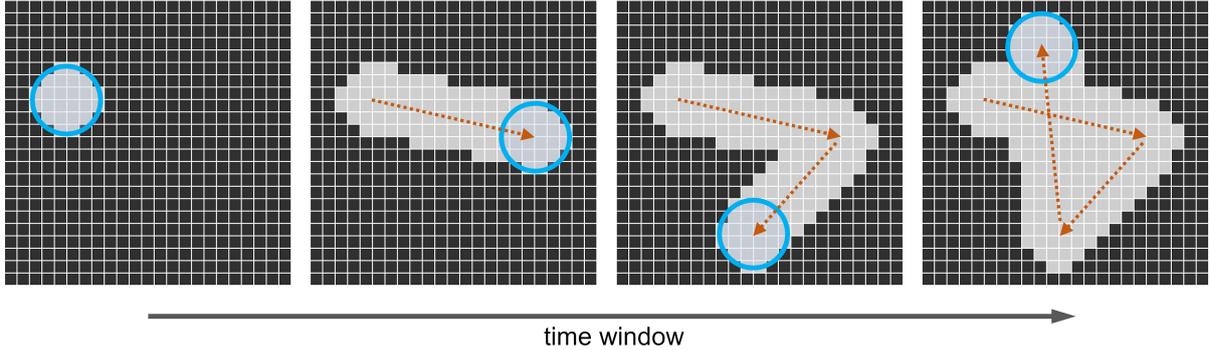}
\caption{Visualization of the concept of the Dynamic Visual Field. The array of black pixels marks a large visual area that could potentially be observed by the participant. The blue circle marks the VF of the patient. Orange arrows indicate gaze movement over a specified time window. Light-grey pixels mark the area that has been observed at any point during the moving time window. The Dynamic Visual Field is defined as the ratio between light-grey pixels and total number of pixels at the end of the time window, meaning in the right-most image.}
\label{fig:DFoVfull}
\end{figure*}

Using the raw eye tracking data, saccades were detected based on an algorithm suggested by Nyström et al. \citep{nyström2010Saccades}. To reduce noise, a moving average over five samples ($\sim$55ms) was applied to the eye tracking data. Then, after applying a low-pass filter of $50^{\circ}/s$ to the data, the mean ($\mu$) and standard deviation ($\sigma$) of the angular velocity of the gaze was calculated for each individual trial. Saccades were registered when gaze velocity surpassed $v_{max} = \mu + 6*\sigma$, with onset and end of the saccade being detected at a threshold of $v_{onset} = \mu + 3*\sigma$.

\begin{itemize}
    \item \textbf{Dynamic Visual Field} The Dynamic Visual Field (DVF) serves as the main indicator for visual performance in this work. It describes the area observed by a participant over a specified shifting time window (Fig. \ref{fig:DFoVfull}) - in case of this study, a three-second time window. The DVF is calculated every 0.5 seconds and averaged at the end of a trial. The gaze direction used to calculate the DVF considers both head movements and eye movements. DVF can be measured in square degree of visual angle, however for better comprehension, the DVF in this work is reported as percentage of a healthy VF at approximately $180^{\circ} * 135^{\circ}$. In other words: a DVF of 10 indicates that a participant observed a total of 10\% of the visual area of a visually healthy subject over the course of the three seconds.

    \item \textbf{Exploratory saccades} Exploratory saccades are defined as saccades that end in a fixation point that lies outside of the area that is visible during the onset of the saccade \citep{ivanov2016eyeMovement}. The parameter is reported as the ratio of exploratory saccades to the total number of saccades within a trial. 

    \item \textbf{Saccade frequency} Saccade frequency describes the number of saccades per second carried out by the participants.

    \item \textbf{Ratio of head-related to eye-related gaze} This parameter describes the ratio by which eye movements and head rotation contribute to the total amplitude of the saccade, respectively. A value of 1 indicates that head movements and eye movements contributed equally to the saccade, a value lower than one indicates that eye movements had a higher contribution, values above 1 indicate higher contribution of head rotation.

    \item \textbf{Ratio of horizontal to vertical gaze movements} Here, the ratio of direction of saccades is reported. A value lower than one indicates that saccades were predominantly horizontally oriented, values higher than one indicate a predominantly vertical orientation of saccades.
\end{itemize}

\subsection{Evaluation and statistical methods} \label{lab:EvaluationProcess}

\subsubsection{Data samples}
Over the course of the study, a total of 6329 Target Tracking trials (3125 from Group A, 3204 from Group B), 6444 Search trials (3205 from Group A, 3239 from Group B), and 5401 Navigation trials (2583 from Group A, 2818 from Group B) have been captured. Due to missing or invalid eye tracking data, some trials had to be excluded from analysis of gaze-related parameters. Trials were excluded from analysis if more than 10\% of eye tracking samples were invalid. Table \ref{tab:invalidGazeData} shows the number and reason of excluded trials. In the remaining trials, the average ratio of invalid eye tracking frames varies between 0.45\% and 0.6\% based on the visual task. Reasons for missing or invalid eye tracking data can include blinking, gaze angles outside of the suggested eye tracking range of 25° from the center of the VF, or temporary hardware failure of the eye tracking device.

\begin{table}[h]
\caption{This table shows the total number of trials per group, the number of trials that had missing eye tracking data, and the number of trials that were excluded from eye tracking analysis due to the ratio of invalid eye tracking frames being above the threshold of 10\%. The values are given for Group A and Group B individually.}\label{tab1}%
\begin{tabular}{p{1.8cm} p{0.5cm} p{0.5cm} p{0.4cm} p{0.4cm} p{0.5cm} p{0.4cm} p{0.4cm}}
\toprule

Visual Task & \multicolumn{2}{|c|}{Trials}  & \multicolumn{2}{|c|}{Missing} & \multicolumn{2}{|c}{Invalid}\\
& A & B & A & B & A & B\\
\midrule
Tracking    & 3125 & 3204 & 5 & 0 & 86 & 4  \\
Search    & 3205 & 3239 & 0 & 2 & 107 & 1  \\
Navigation    & 2583   & 2818  & 42 & 0 & 143 & 3 \\
\botrule
\label{tab:invalidGazeData}
\end{tabular}
\end{table}

Outliers in the dataset have been adjusted by limiting their values to within three times the standard deviation from the mean \citep{Sullivan2021Outliers}.

\subsubsection{Statistical analysis}
The data analysis in this study was conducted using the R programming language, facilitated by the RStudio graphical interface with the 'nlme' and 'lme4' libraries. The primary objective of this study was to evaluate the similarity of parameters between two groups. Traditional statistical tests like Anova or Linear Mixed Models, while effective at detecting significant differences between groups, fall short in demonstrating equivalence between them. Instances where these models fail to reject the null hypothesis only describe the absence of significant effect, which does not necessarily confirm the equivalence between the groups. To address this, we employed a statistical approach known as 'two one-sided tests for equivalence' (TOST) \citep{Schuirmann1987TOST}. The TOST method involves two one-sided hypothesis tests. The first test assesses whether the mean difference between Group A and Group B is greater than a pre-defined delta value (upper equivalence limit). The second test assesses whether the mean difference is less than negative delta (lower equivalence limit). If both one-sided tests fail to reject their respective null hypotheses, it provides evidence that the two groups are within the specified delta range.

Each analysis starts by fitting a linear regression model to the data, using the respective measurement parameter as dependent variable and the parameters 'training session' and 'trial number' as fixed effects. By including these factors, the model minimizes the effects of learning-related effects from its residuals, leaving mainly participant-dependent effects as well as stochastic noise. Next, a delta value for the TOST has to be defined. Following an approach by Ng et al. \citep{ng2001ChoiceOfDelta}, the delta value $\delta$ is set to 0.2 times the standard deviation of the samples. Confidence level is set to 0.95 following statistical conventions. With all parameters in place, the TOST is conducted for each investigated parameter on each of the three visual task results.

When the TOST fails to establish significant equivalence between two groups for a given parameter, it prompts the question about whether there is a significant effect between the two groups or if the results are inconclusive. The latter suggests that the sample size does not provide sufficient statistical power to draw clear conclusions about the relation between the two groups regarding the respective parameter. To test for significant effects between the groups, Linear Mixed Models (LMMs) are applied for all continuous parameters. An exception is the 'Number of Collisions' parameter in the navigation task, which has discrete values and exhibits a high zero-inflation. Here, a negative-binomial Generalized Linear Mixed Model (nbGLMM) was employed, as it is particularly suitable for such data types \citep{bates2015LME4}. Additionally, the 'Trial duration' parameter of the navigation task was transformed using a logarithmic function to improve normal distribution of samples. The general structure of the LMMs was comprised of the same components. The respective measurement parameter was included as dependent variable. Fixed effects included Trial Number as well as the interaction term between Training Session and Group. The interaction term allows to test whether the average learning rate of the two groups significantly differs. Lastly, the individual participant parameter was included as random factor, taking into account both random intercepts and random slopes. This approach recognizes that each participant has unique inherent performance levels and gaze characteristics (random intercept), as well as distinct learning patterns over the course of the study (random slope). To ensure the appropriateness of these models, the normal distribution of residuals was assessed using QQ-plots as a visual indicator.

Finally, the relationship between task performance and VF size is investigated using LMMs. For these, the performance parameters were used as dependent variable, Training Session and VF size are included as interacting fixed factors, and the individual participant parameter was again included as random factor. Notably, only a random intercept was taken into account for the random factor due to the models failing to converge when considering a random slope.

Statistical evaluation of the questionnaires is not feasible due to the low number of samples. Results are reported as raw data.

\section{Results}\label{lab:results}

\begin{figure*}[htbp]%
\centering
\includegraphics[width=\linewidth]{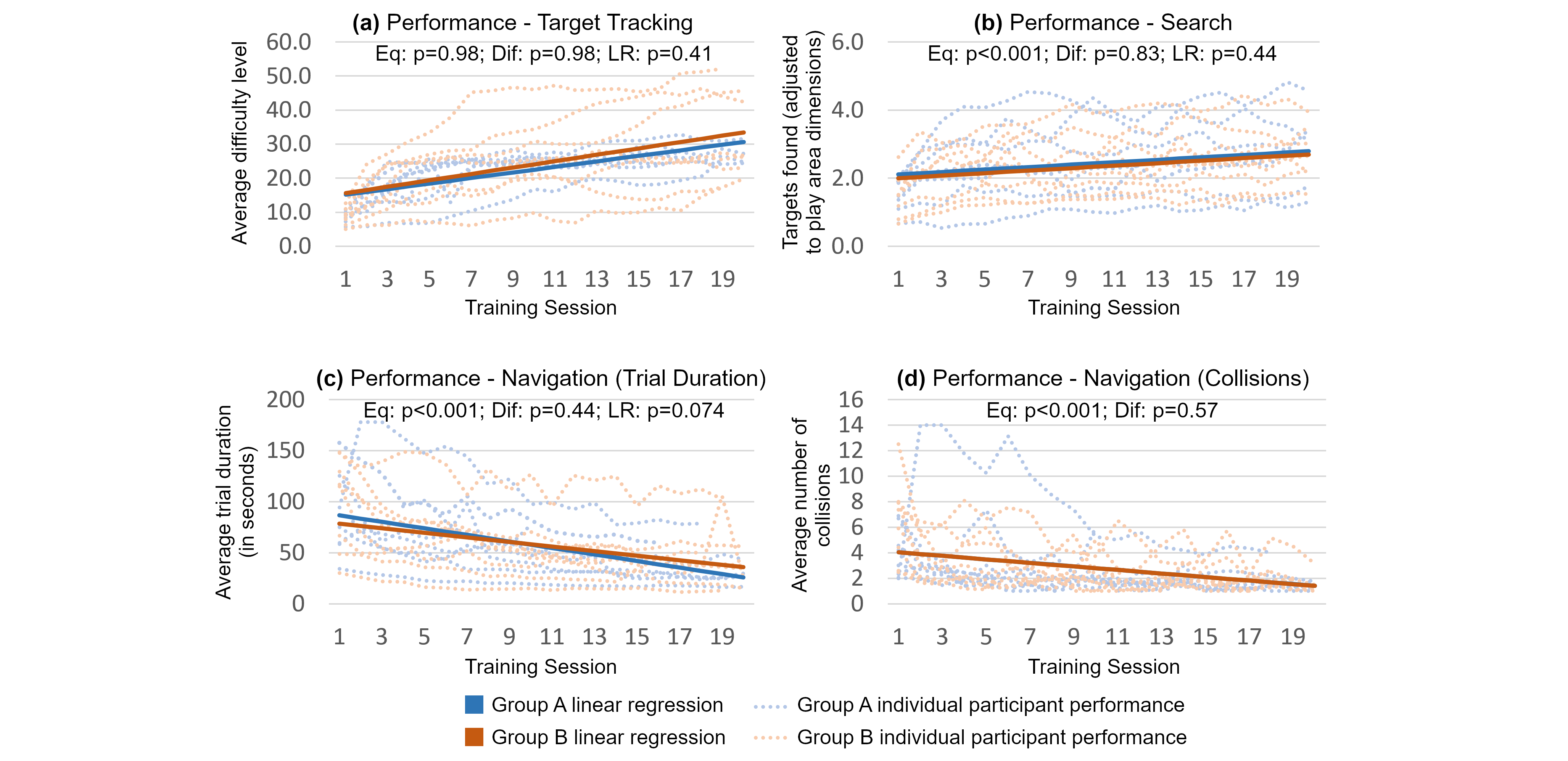}
\caption{Results for the task performance in the three visual tasks. Dotted lines show the average results for each individual participant, continuous lines show the average regression of all trials. Results for trial duration and number of collisions in the navigation task are shown separately. Non-logarithmic trial duration values are shown for better visualization despite statistical evaluation only considering the logarithm of trial duration for better normal distribution. Each plot includes the resulting p-values of the statistical models conducted: Eq describes the equivalence between groups found with TOST, Dif describes the effect between groups found with an LMM, and LR describes the interaction effect between the Training Session and Group parameter, meaning differences in the learning rate between groups. Interaction terms are not feasible in the nbGLMM, thus no p-value for LR is reported under (d).}
\label{fig:PerformanceReference}
\end{figure*}

\subsection{Equivalence and statistical effects between groups}
In this first section, we will compare the results of Group A and Group B with each other across all measurement parameters and for all three visual tasks. The objective is to assess the extent to which the simulation of tunnel vision in visually healthy participants captures various aspects of the condition. For each condition, three statistical findings are reported:
\begin{enumerate}
    \item Equivalence (Eq): This indicates the p-value obtained from the TOST analysis. A p-value $<0.05$ signifies significant equivalence between both groups, implying that their results in the given condition are comparable.

    \item Difference (Dif): This reflects the p-value derived from the LMM or nbGLMM, as applicable. A p-value $<0.05$ suggests significant differences between the results of Group A and Group B for the specific condition.

    \item Learning Rate (LR): This assesses whether the two groups exhibit significantly different effects in relation to the dependent variable and the Training Session parameter. This comparison is evaluated within the same LMM or nbGLMM models as before. In this context, a p-value $<0.05$ indicates that the two groups display divergent changes in average performance or gaze behavior over the duration of the study, which can be interpreted as one group improving or adapting faster than the other.
\end{enumerate}

\subsubsection{Performance}
Fig. \ref{fig:PerformanceReference} shows the performance displayed by both groups in the different tasks over the course of training. Raw data of the results is found in the supplementary files (File S1)

Both groups show significant improvements in performance over the course of the study in all visual tasks ($p<0.001$ for all four conditions). For equivalence, it is found that in the target tracking task, null hypothesis of statistical difference is not rejected ($p=0.98$), indicating that the results of the two groups are not equivalent. For the three other parameters, however, null hypothesis of statistical difference is rejected ($p<0.001$ for all three parameters), thus the performance of both groups can be assumed equivalent in both the search task and the navigation task. As a second step, the parameters were analyzed regarding statistically significant differences using an LMM or nbGLMM, respectively. However, no significant differences between Group A and Group B were found for any of the performance parameters, both regarding the mean of results as well as learning rates.

\subsubsection{Gaze characteristics}
\begin{figure*}[htbp]%
\centering
\includegraphics[width=\linewidth]{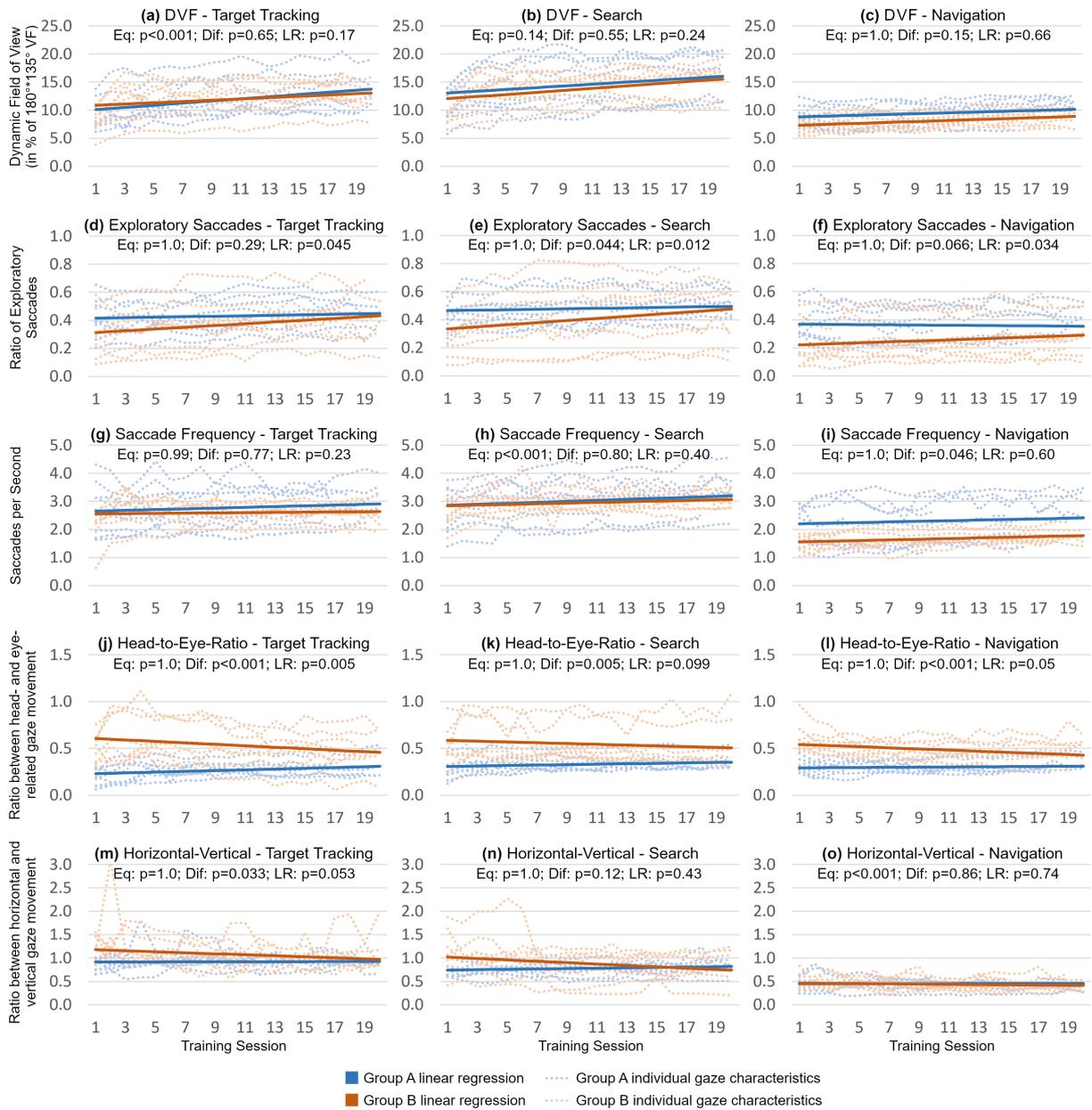}
\caption{Results of the gaze characteristic parameters in the three different visual tasks, showing individual results of participants and average regression for Group A and B. Each plot includes the resulting p-values of the statistical models conducted: Eq describes the equivalence between groups found with TOST, Dif describes the effect between groups found with a LMM, and LR describes the interaction effect between the Training Session and Group parameter, meaning differences in the learning rate between groups. Head-to-Eye-Ratio describes the amount of head movement (in visual angles) divided by the amount of eye movement contributing to saccades. Horizontal-to-Vertical describes the amount of gaze movement on the elevation axis divided by the amount of gaze movement on the azimuth axis.}
\label{fig:DFoVComparison}
\end{figure*}

The results of the five gaze characteristic parameters tested within this work are displayed in Fig. \ref{fig:DFoVComparison}. Information about statistical significance of both equivalence and difference between results of both groups is noted within the figure. The null hypothesis for statistical difference is rejected in three conditions: the DVF in the target tracking task, the saccade frequency in the search task, and the ratio between horizontal and vertical gaze direction in the navigation task. As for conditions where significant effects are found between the results of Group A and B, these include the ratio of exploratory saccades in the search task, the saccade frequency in the navigation task, as well as the ratio of head-related gaze movements to eye-related gaze movements in all three tasks. In the other condition, neither significant equivalence nor difference can be concluded from the statistical tests. Regarding differences in the learning rate between the groups, noticeable parameters are the ratio of exploratory saccades as well as the head-to-eye ratio, both of which show a convergence between groups over the course of the study in almost all tasks.

\subsection{Influence of visual field size}

\begin{figure}[hbp]%
\centering
\includegraphics[width=\columnwidth]{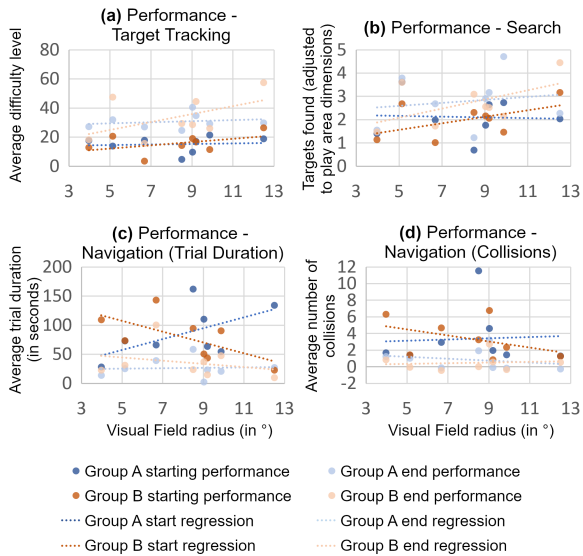}
\caption{Relationship between the different task performances and the average radius of the - real or simulated - VFs. The displayed results show the approximate performance at the start and end of the study, respectively.}
\label{fig:VFinfluence}
\end{figure}

In this section, it is investigated how the size of the VF of patients from Group A - and respectively the size of the simulated tunnel vision of the matched participant in Group B - affects the performance. Fig. \ref{fig:VFinfluence} shows the relationships between these parameters. The performance values displayed in these graphs are based on the predicted performance for the 1\textsuperscript{st} and 20\textsuperscript{th} training session, calculated using the results of linear regression. While some trends are visible in the graphs, no statistically significant effects between VF radius and task performance were found for any of the visual tasks. 

\subsection{Questionnaires}
\begin{figure*}[htbp]%
\centering
\includegraphics[width=\linewidth]{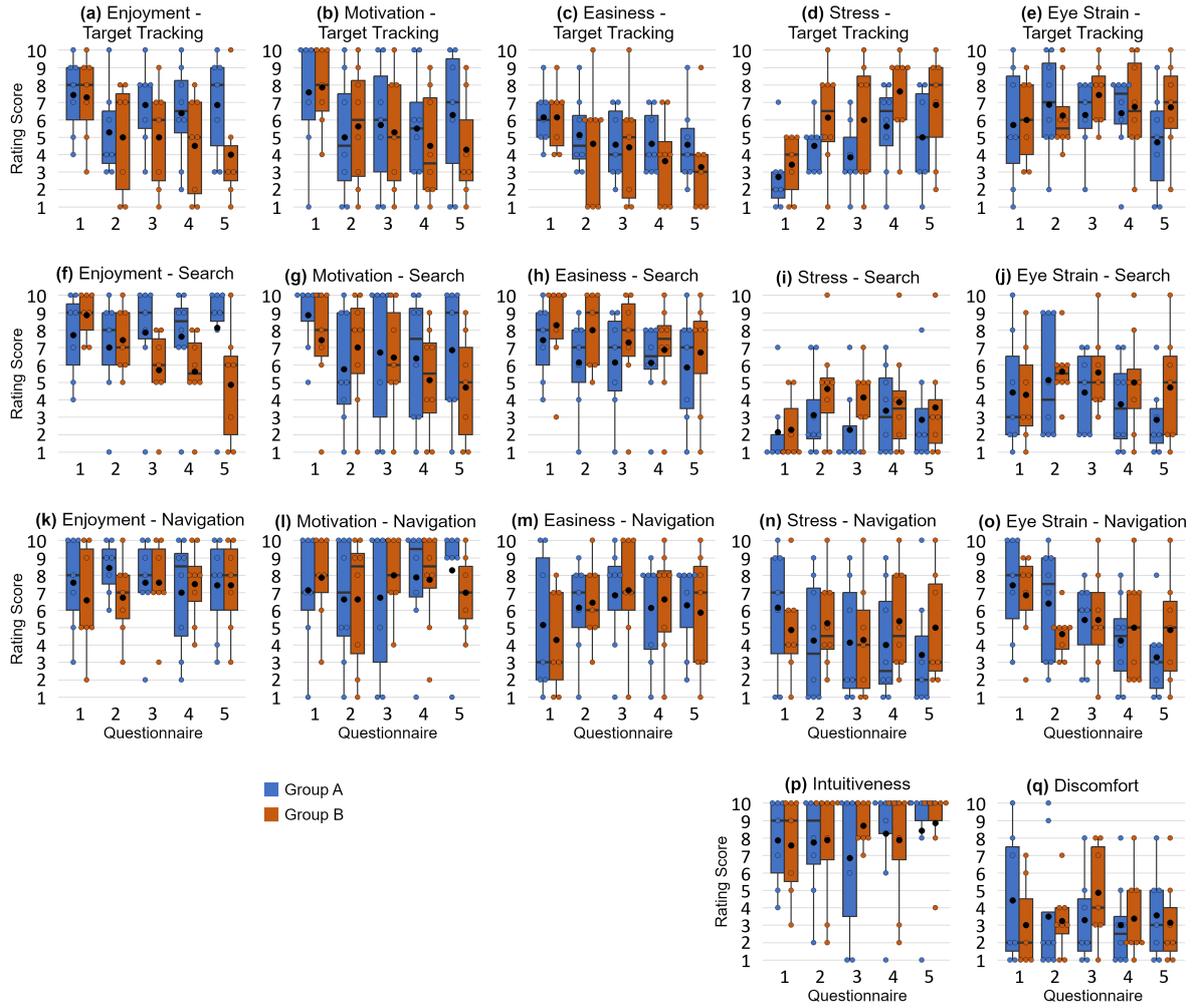}
\caption{Rating results for different aspects of the visual training. Questionnaires 1-5 were filled after 1, 5, 10, 15 and 20 training sessions. A rating of 10 equals 'very high', a rating of 1 equals 'very low'.}
\label{fig:QuestionnaireResults}
\end{figure*}

The questionnaires that were filled by both groups after the course of the study give insight into the qualitative results for the gaze training. The average rankings as well as the distribution of individual scores are shown in Fig. \ref{fig:QuestionnaireResults}. The results show a high variance between different participants even within the same group and condition, indicating high influence of subjective perception and preference. Despite this, it is noticeable that Group A's rating is generally higher regarding task enjoyment and motivation, and lower regarding stress and eye strain. The trend is especially noticeable towards the end of the study. No clearly noticeable differences between groups are found in the ratings regarding easiness, intuitiveness, and discomfort.

\section{Discussion}
In this study, we have investigated the degree by which the simulation of visual field defects - specifically the simulation of tunnel vision scotoma - in visually healthy participants within a virtual environment can represent the effects of real visual field defects. We have analyzed and compared the results of two age-matched groups consisting of eight Retinitis pigmentosa patients and eight visually healthy participants, respectively, in three visual tasks over the course of four weeks. The findings allow to draw conclusions about which aspects of vision and vision-related task performance can feasibly be reproduced through VR-based simulation, which aspects show clear differences, and how this relation changes over the duration of the study.

Our findings suggest that the simulation of tunnel vision in visually healthy participants is well suited to accurately represent performance within different vision-based VR tasks. Three out of four investigated performance parameters (Fig. \ref{fig:PerformanceReference} b-d) - relating to tasks of static target search and navigation - show significant equivalence between RP patients (Group A) and participants with simulated tunnel vision (Group B). For the fourth parameter (Fig. \ref{fig:PerformanceReference} a) - related to the task of moving target tracking -, equivalence between the two data sets can neither be assumed nor rejected based on statistical results. 

Differences between the two groups' results arise when evaluating specific characteristics of the gaze behavior of participants (Fig. \ref{fig:DFoVComparison}). Namely, we have investigated (i) the Dynamic Visual Field (DVF), which describes the amount and efficiency of gaze movements for observing the visual surroundings; (ii) the ratio of exploratory saccades compared to the total number of saccades within a trial; (iii) the frequency of saccades; (iv) the ratio between head movements and eye movements that contribute to overall gaze; (v) the ratio between vertical and horizontal direction of gaze movements. Here, only few results and only in individual visual tasks are found significantly equivalent between groups. Given the large number of individual analyses conducted regarding the different gaze characteristics, it is possible that the findings are influenced by the Multiple Comparisons Problem \citep{sullivan2021MultipleComparisons}. Thus, it is more feasible to focus only on those parameters where notable features occur in multiple related conditions. The first of these is the ratio of exploratory saccades (Fig. \ref{fig:DFoVComparison} d-f), which is found to be significantly higher in Group A in the search task, with visual trends in the target tracking task and navigation task supporting this.

Next, it can be observed that in all three tasks, Group B displayed a significantly higher ratio of head movements to eye movements compared to Group A (Fig. \ref{fig:DFoVComparison} j-l). This behavior is likely related to the technical implementation of the tunnel vision simulation. As was stated in section \ref{lab:specs}, Stein et al. \citep{Stein2021Latency} found the end-to-end latency of VR devices applying the tobii eye tracking system to be around 79ms. Albert et al. \citep{Albert2017EyeTrackingLatency} report that for foveated rendering - a technique similarly dependent on low eye-tracking latency - the range of latency at which participants start reporting a noticeable delay is between 50-70ms. It can thus be assumed that the simulation of tunnel vision applied in this work had an at least subconsciously noticeable delay. Notably, one participant from Group B did mention perceiving a slight delay in eye tracking, but stated that it was not perceived as obstructive. While no exact information about the latency between head-related movement and the displayed visual content - also called Motion-to-Photon (M2P) latency - is given for the Pico Neo 2 Eye, these measurements exist for similar commercial VR devices \citep{Optopfidelity2020HeadTrackingLatency} and were found to range between 0 and 5ms. Assuming the Pico Neo 2 Eye's M2P latency is comparable to these values, it would be substantially lower than the eye-tracking latency. This difference could explain the significantly higher ratio of head movements observed in Group B, as lower latency incentivizes higher reliance on the respective type of tracking. The risk of eye-tracking latency influencing the results was known before experimental trials commenced. However, at the time, the Pico Neo 2 Eye was the only commercially available stand-alone VR device that provided built-in eye-tracking. Alternatives with lower end-to-end latencies, such as the Fove-0 (45ms) or Varjo VR-1 (57ms) \citep{Stein2021Latency}, would require complex setup, including connection to external processing and tracking devices, making them not feasible for use in unsupervised at-home training. 

Lastly, it is noticeable that for most gaze characteristics in which both groups show significantly different results - which mostly includes the parameters of exploratory saccade ratio (Fig. \ref{fig:DFoVComparison} d-f) and head movements to eye movements (Fig. \ref{fig:DFoVComparison} j-l) -, the results of the two groups show significant trends to converge over the course of the study. This suggests that even significant differences in those gaze characteristics may be minimized through extended adaption to the simulated visual field defects. It remains to be tested whether this effect is transferable between tasks, meaning whether sighted participants could be 'trained' to adapt gaze characteristics of patients in one task and then display similar performance and gaze characteristics in different tasks.

Regarding the influence of VF size on performance, as shown in Fig. \ref{fig:VFinfluence}, visual trends seem to suggest that the performance of Group B is generally more negatively affected by smaller VFs. Given the lack of statistical significance of these trends, however, no validated statements can be made regarding these effects.

In the questionnaires, it is overall noticeable that the ratings in both groups show strong variations between individual participants, oftentimes covering almost the entire range from lowest to highest possible score rating. This shows that the perception of the tasks is highly subjective and difficult to predict. On average, it appears as though RP patients report higher enjoyment and lower perceived stress during the visual tasks, especially in the Target Tracking and search tasks. However, given the high variance between answers, it is once again not feasible to derive solid conclusions from the questionnaire results.  

Different previous studies have shown that the applied approaches to simulate visual field impairments have the desired effects on visually healthy participants in that they reduce their performance in visual tasks \citep{jones2020Perspectives, krosl2019VRCataractSimulation, vayrynen2016VRCitySimulation}. However, to the best of our knowledge, no previous study has quantitatively assessed performance differences and differences in gaze behavior between real patients and participants with simulated visual field impairments. This makes a comparison between results of different studies difficult and not feasible.

While the results of this work can be relevant for future vision impairment studies and accessibility test setups utilizing VR, some limitations of the findings have to be considered. For one, the experimental setup was designed to assess the influence of visual field defect only. Other aspects of visual impairments, such as limited visual acuity or contrast sensitivity, glare sensitivity, or night blindness, were not considered. The experimental environment was designed to minimize any influence by ensuring that all patients were able to effortlessly detect and recognize any visual target or element within their VF. Furthermore, all measured results of this study were exclusively gathered within a virtual environment, with participants maintaining a stationary seated position. Consequently, it is important to note that these results hold no direct relevance for the practical application of simulations in real-world scenarios. This limitation might initially appear redundant, given that the utilization of a VR headset is essential for the accurate simulation of gaze-contingent visual field defects. However, technology such as mixed reality, while not yet widely popular and commercially available, could change this in the near future. Mixed Reality headsets function similar to VR headsets in that they display a scene to the user through head-mounted displays. However, Mixed Reality devices have the ability to capture the real world through front-facing cameras, projecting the images to the device's displays to allow the user an almost-natural view of the surrounding area. However, as this image of the real world is digitally projected to the screens, the displayed scene can be freely modified using additional virtual contents that are overlayed with the real-world capture. Such virtual content could be a visual field defect mask, which essentially allows to apply the simulation used in this work to real-world tasks. The evaluation of this technology, its feasibility for real-world vision impairment simulation, and the comparison of its results with those of real patients offer interesting questions for future research.

Based on our findings, expanding a group of patients living with visual field defects with a larger cohort of visually healthy participants with simulated visual field defect seems largely feasible if the following conditions are met: (i) The study or accessibility test for which the cohort is recruited can provide meaningful results if realized within a virtual environment; (ii) The study or test mainly focuses on quantitative rather than qualitative results; (iii) The primary emphasis of the study or test is on evaluating task performance rather than specific gaze behavior; (iv) The study or test is primarily concerned with the average results of a group, rather than effects that depend on characteristics of the individual participant; (v) Ideally, the visually healthy participants have the possibility to adapt to the simulation for several hours before taking part in the actual assessment.

\section{Conclusion}
We evaluated whether the simulation of tunnel vision in visually healthy subjects within a virtual reality setup can accurately mirror performance and gaze behavior of actual patients living with the condition. Findings suggest that the group with simulated tunnel vision succeeds in accurately representing the task performance of the patient group. Gaze characteristics are largely not found to be significantly equivalent and sometimes even significantly differ between groups. However, over the course of 10 hours of visual task execution, several gaze characteristics between the two groups begin to converge, implying that the extent to which the simulation can represent the actual visual impairment increases with expanded use of the simulation. 

\section{Supplementary files}
\begin{itemize}
    \item \textbf{File S1} List of raw data for all recorded trials.

    \item \textbf{Video S1} Video showcase of the three visual tasks, both with and without simulated tunnel vision.
\end{itemize}

\subsection*{Acknowledgements} We want to thank Enkelejda Kasneci for her indispensable guidance and support for this work. Furthermore, we thank our participants who made this study possible.
\subsection*{Competing interests} This work was done in an Industry-on-Campus cooperation between the University of Tübingen and Carl Zeiss Vision International GmbH. Three of the authors, Nora Castner, Iliya Ivanov, and Siegfried Wahl, are employees of Carl Zeiss Vision International GmbH. Their affiliation with Carl Zeiss Vision International GmbH had no influence in the study. There are no competing interests related to employment, consultancy, patents, products in development, or marketed products.
\subsection*{Funding} This work was supported by the Deutsche Forschungsgemeinschaft (DFG), URL: https://www.dfg.de/en/ (Grant: DFG IV 167/5-1 to II). The funder did not play any role in study design, data collection and analysis, decision to publish, or preparation of the manuscript.
\subsection*{Author contributions} Conceptualization: Alexander Neugebauer, Iliya Ivanov, Siegfried Wahl; Methodology: Alexander Neugebauer; Medical guidance and supervision: Katarina Stingl; Formal analysis and investigation: Alexander Neugebauer, Nora Castner, Björn Severitt; Writing - original draft preparation: Alexander Neugebauer; Writing - review and editing: All Authors; Funding acquisition: Iliya Ivanov; Supervision: Siegfried Wahl.

\clearpage

\begin{appendices}

\section{Visual fields}\label{secA1}

\noindent
\begin{minipage}{1.0\textwidth}
\centering
\includegraphics[width=\linewidth]{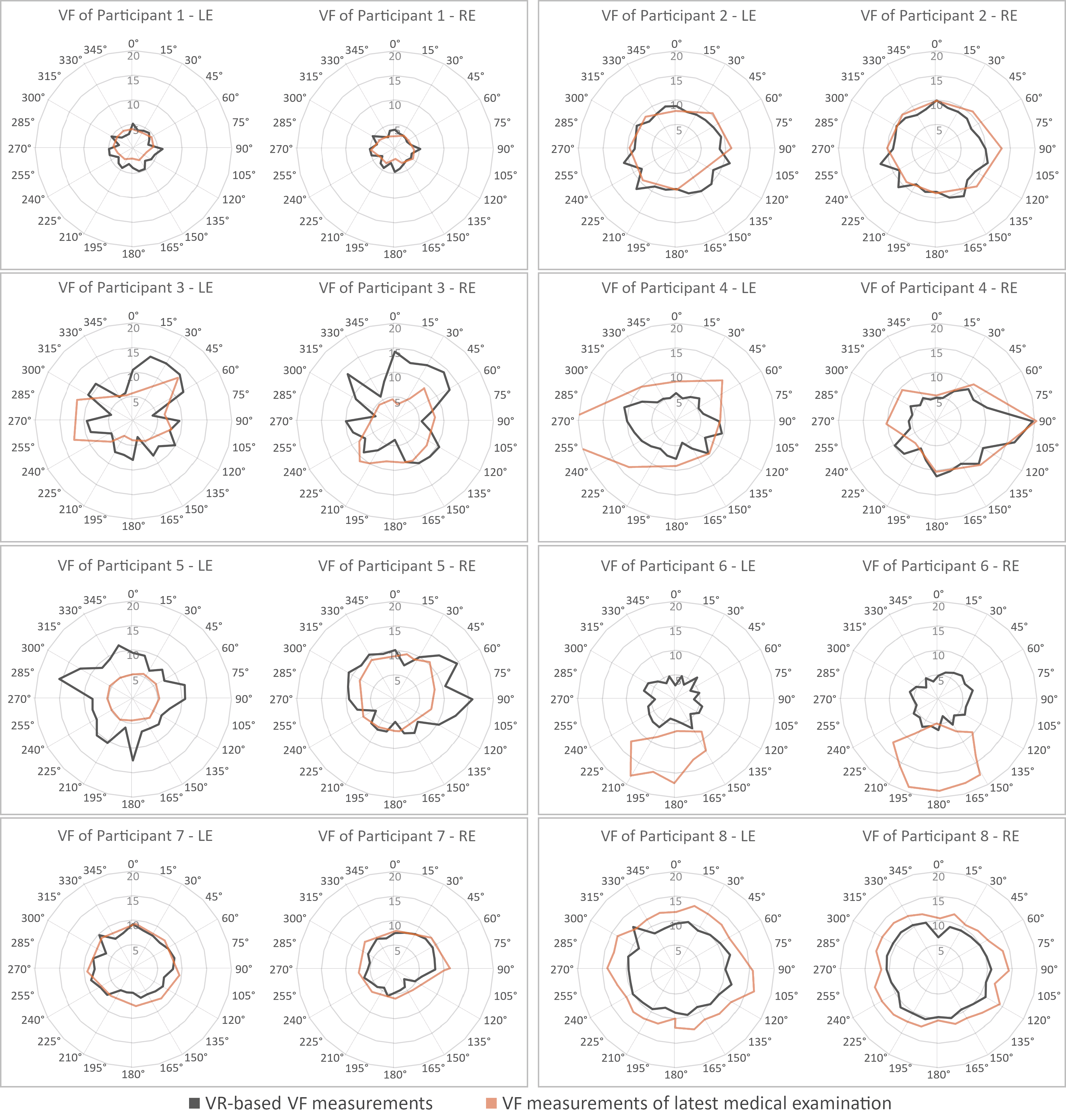}
\captionof{figure}{Comparison of the VFs of patients from their most recent medical exam and the VFs directly measured within the Virtual Reality setup.}
\label{fig:VisualFieldComparison}
\end{minipage}

\FloatBarrier

Fig. \ref{fig:VisualFieldComparison} displays the visual fields of the eight RP patients of group A, as reported in \citep{neugebauer2023gazeTraining}.

\end{appendices}

\clearpage

\bibliography{sn-bibliography}

\end{document}